\documentclass[twoside,twocolumn,english,aps,prx]{revtex4-1}

  \usepackage[usenames,dvipsnames]{xcolor}
	\usepackage{amsmath}
	\usepackage{amsfonts}
	\usepackage{amssymb}
	\usepackage[colorlinks=true,citecolor=blue,linkcolor=red]{hyperref}
	\usepackage{graphicx}
	\usepackage{bbold}					
	\usepackage[makeroom]{cancel}		
	\usepackage{multirow}				
	\usepackage[normalem]{ulem}        
	\usepackage{array}
	\usepackage{pdfpages}

	\makeatletter
	\AtBeginDocument{\let\LS@rot\@undefined}
	\makeatother
	
	\newcolumntype{x}[1]{>{\centering\let\newline\\\arraybackslash\hspace{0pt}}p{#1}}
	\renewcommand{\Re}{\operatorname{Re}}
	
	\DeclareMathOperator{\tr}{tr}  		

\begin{document}
\title{Distinguishing quantum dynamics via Markovianity
and Non-Markovianity}

\author{Yi Zuo$^{1,2}$}
\author{Qinghong Yang$^{3}$}\email[Corresponding to: ]{yqh19@tsinghua.org.cn}
\author{Bang-Gui Liu$^{1,2}$}\email[Corresponding to: ]{bgliu@iphy.ac.cn}

\affiliation{$^{1}$Beijing National Laboratory for Condensed Matter Physics and Institute of Physics,Chinese Academy of Sciences, Beijing 100190, China}
\affiliation{$^{2}$University of Chinese Academy of Sciences, Beijing 100049, China}
\affiliation{$^{3}$State Key Laboratory of Low Dimensional Quantum Physics, Department of Physics, Tsinghua University, Beijing, 100084, China}
\date{\today}

\begin{abstract}
To study various quantum dynamics, it is important to develop effective methods to detect and distinguish different quantum dynamics. A common non-demolition approach is to couple an auxiliary system (ancilla) to the target system, and to measure the ancilla only. By doing so, the target system becomes an environment for the ancilla. Thus, different quantum dynamics of target systems will correspond to different environment properties. Here, we analytically study $\mathrm{XX}$ spin chains presenting different kinds of quantum dynamics, namely localized, delocalized, and dephasing dynamics, and build connections between Markovianity and non-Markovianity --- the two most common properties of an environment. For a qubit coupled to the $\mathrm{XX}$ chain, we derived the reduced density matrix of the qubit through the projection method. Furthermore, when dephasing noise was introduced to the $\mathrm{XX}$ chain, we generalized the projection method by introducing an {\em open-system interaction picture} --- a modification of the Dirac interaction picture. By calculating the reduced density matrix for the qubit analytically and numerically, we found that the delocalized (localized) chain corresponds to the Markovian (non-Markovian) bath when boundary effects are not considered, and the feature of the chain with dephasing noise as a bath is dependent on the dephasing strength. The three kinds of quantum dynamics can be distinguished by measuring the qubit only.
\end{abstract}


\maketitle
\section{Introduction}

In closed quantum systems, quantum dynamics is governed by the Schrödinger equation, leading to unitary evolution and the preservation of quantum coherence. Despite this, closed quantum systems do not necessarily exhibit ergodicity, and different types of dynamics can emerge. For example, thermalization and localization represent two distinct behaviors: in thermalizing systems, a subsystem loses information to the rest of the system and reaches effective thermal equilibrium \cite{Abanin}; in contrast, localized systems retain memory of their initial states even after long-time evolution \cite{Hamacher:2002tk,Banuls:2017wf}, supporting phases such as Anderson localization \cite{Anderson:1958uc} and many-body localization (MBL) \cite{Gornyi2005,BASKO20061126}.

When quantum systems are open, interactions with external environments give rise to new dynamical features. Decoherence, for instance, suppresses quantum coherence and alters transport properties, driving systems from ballistic to diffusive dynamics \cite{Li,Medvedyeva:2016wh,Mendoza-Arenas:2013ul}. Under strong decoherence, an open system eventually approaches a maximally mixed state. Although global information is lost, the timescale for local information decay differs from that of thermalization.

Recent advances in experimental platforms—including cold atomic gases \cite{RevModPhys.80.885}, nitrogen-vacancy centers \cite{schirhagl2014nitrogen}, superconducting circuits \cite{guo2021observation}, and trapped ions \cite{blatt2012quantum}—have enabled the preparation and probing of complex quantum dynamics. Methods such as local observable measurements \cite{jurcevic2017direct}, out-of-time-ordered correlators (OTOCs) \cite{garttner2017measuring,Kukuljan:2017tq,campisi2017thermodynamics,Shen:2017vx}, and Loschmidt echo experiments \cite{braumuller2022probing} have been developed to study these dynamics. However, extracting full information often requires significant resources, including multiple copies of initial states and repeated measurements, with resource costs scaling unfavorably with system size.

To reduce measurement overhead, indirect measurement schemes have been proposed, where an auxiliary system is coupled to the target system and only the auxiliary system is measured. For instance, coupling a single qubit to a quantum chain allows the qubit’s decoherence dynamics to serve as a probe for the target system’s properties \cite{Nieuwenburg2016uq,Vasseur2015,Eleuch:2017vs,gaikwad2024entanglement}. In this context, the quantum chain effectively acts as an environment for the auxiliary qubit, and the dynamical properties of the chain manifest in the probe's evolution.

From the perspective of open quantum systems, environments are typically classified as Markovian or non-Markovian \cite{lindblad1979non,laine2010measure,Breuer2016}. A Markovian environment continuously dissipates information from the system, while a non-Markovian environment retains information and can feed it back. Thus, analyzing the non-Markovianity of a probe qubit’s dynamics provides a window into the properties of its surrounding environment.

Motivated by this, our goal is to distinguish different types of quantum dynamics in the target system by analyzing the Markovian or non-Markovian features of a coupled probe qubit. This allows for efficient characterization of the target system with reduced experimental resources, and sheds light on the relationship between system dynamics and environmental memory effects.

The dynamics of a quantum probe is typically modeled by deriving a master equation for its reduced density matrix, starting from the von Neumann equation for the total system. Standard techniques such as the Nakajima-Zwanzig (NZ) \cite{Nakajima1958,Zwanzig1960} and time-convolutionless (TCL) \cite{ChaturvediShibata1979} master equations are widely used under the assumption that the total system is closed. However, in practical settings such as quantum computing platforms, the target system itself is subject to external noise that cannot be fully captured by a closed-system Hamiltonian description. Consequently, the total system comprising the probe and target becomes an open system, and standard master equation approaches become inapplicable.

This work addresses this more general scenario. We propose a theoretical framework suitable for open total systems by introducing the open-system interaction picture, a generalization of the Dirac interaction picture, which allows us to derive effective dynamical equations even when the total system is not closed. Within this framework, we extend the conventional TCL master equation and formulate the Open Time-Convolutionless (Open-TCL) approach.

\begin{figure}[t]
\centering
\includegraphics[height=3.0 cm]{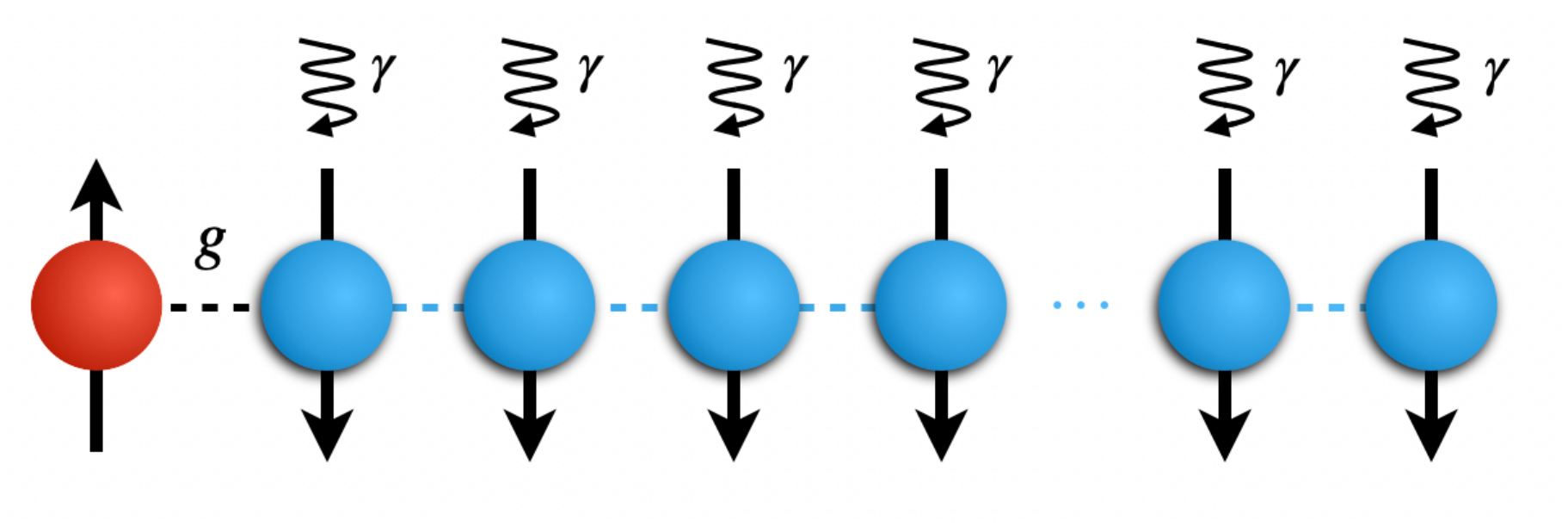}
\caption{A single qubit S (the red one) coupling to a XX spin chain with coupling strength $g$, and the XX chain can be exposed to some dephasing noise, with the dephasing strength  $\gamma$. When $\gamma=0$, the total system is a closed quantum system.} 
\label{Fig:setup}
\end{figure}

As a concrete application, we consider a model consisting of a quantum spin-XX chain (the target system) coupled at its left end to a probe qubit \( S \). We analyze three dynamical regimes of the spin chain: (i) delocalized dynamics without decoherence noise, (ii) localized dynamics without decoherence, and (iii) diffusive dynamics induced by strong dephasing noise \cite{Li,Medvedyeva:2016wh,Mendoza-Arenas:2013ul}. The model setup is illustrated in Fig.~\ref{Fig:setup}. By measuring the probability of the probe qubit \( S \) being in the spin-up state, we extract information about the environment’s dynamical properties and investigate their relation to Markovianity and non-Markovianity.

This paper is organized as follows. In Section~\ref{sec21}, we review the projection operator method in the standard Dirac interaction picture. In Section~\ref{sec22}, we develop the Open Time-Convolutionless (Open-TCL) approach by extending the projection method to open-system settings. In Section~\ref{secxx}, we present the specific model and apply the Open-TCL framework to analyze its dynamical behavior under different conditions. Finally, Section~\ref{sec4} discusses the results and concludes the paper.

\section{Projection Operator Method in the Dirac Interaction Picture}\label{sec21}

In this section, we briefly review the time-convolutionless (TCL) projection operator method formulated within the Dirac interaction picture, which provides the theoretical foundation for our subsequent development. The projection operator technique systematically derives effective equations of motion for a subsystem by eliminating environmental degrees of freedom. It leads to master equations such as the Nakajima-Zwanzig (NZ) equation and the time-convolutionless (TCL) master equation, widely used to describe open quantum systems~\cite{Breuer2002}.

For a general system $\mathrm{S}$ coupled to a bath $
\mathrm{B}$, when the total system is closed, that is, the system $\mathrm{S}$ and bath $\mathrm{B}$ only couple to each other without other external couplings, the evolution of the combined system is governed by the total Hamiltonian $H=H_s+H_b+H_i$, where $H_s$ is the Hamiltonian of the system $\mathrm{S}$, $H_b$ is the Hamiltonian of the bath $\mathrm{B}$, and $H_i$ is the interaction between $\mathrm{S}$ and $\mathrm{B}$.
In this scenario, one can study the evolution perturbatively in the Dirac interaction picture. Then, the von Neumann equation
for the total density matrix becomes 
 \begin{equation}\label{eq3}
\begin{aligned}
 \frac{\partial}{\partial t}\rho(t)=-ig\left[H_{i}(t),\rho(t)\right]\equiv g\mathcal{\mathcal{L}}(t)\rho(t),
\end{aligned}
\end{equation}
Here, $\rho(t)$ is the density matrix of the total system, including both the system and the environment. The interaction Hamiltonian in the Dirac interaction picture is given by $H_i(t) = e^{iH_0 t} H_i e^{-iH_0 t}$, where $H_0 = H_s + H_b$ is the free Hamiltonian of the system and the bath. The Liouville superoperator $\mathcal{L}(t)$ captures the generator of time evolution in the interaction picture and acts on operators via the commutator with $H_i(t)$.

To obtain a dynamical equation for the reduced density matrix of the system $\mathrm{S}$, we briefly introduce the Time-Convolutionless (TCL) projection operator technique, which effectively describes the dynamics of the open system S. The fundamental idea of the TCL approach is to consider the process of tracing out the environment as a projection in the total system's state space. The projection is realized by a superoperator $\mathfrak{\mathcal{P}}$, which operates according to  $\mathfrak{\mathcal{P}}\rho(t) = \text{tr}_{\mathrm{b}}[\rho(t)] \otimes \rho_{b}$
, and $\rho_{b}$ is some fixed state of the bath $\mathrm{B}$.  The term $\mathfrak{\mathcal{P}}\rho(t)$ is the relevant part  of the total system's density matrix $\rho$, whereas the complementary part is then defined as $\mathcal{Q}\rho(t) \equiv \rho(t) - \mathfrak{\mathcal{P}}\rho(t)$. Finally, the exact dynamical equation for the relevant part   $\mathfrak{\mathcal{P}}\rho(t) $ can be expressed as
 \begin{equation}\label{eq:tclfome}
\begin{aligned}
\frac{\partial}{\partial t}\mathfrak{\mathcal{P}}\rho(t)=\mathcal{K}(t)\mathfrak{\mathcal{P}}\rho(t)+\mathcal{I}(t)\mathcal{Q}\rho(t_{0}),
\end{aligned}
\end{equation}
which is the exact time-convolutionless form of the master equation
.  Here, $\mathcal{K}(t)$ can be written as 
\begin{equation}\label{eq41}
\begin{aligned}
\mathcal{K}(t) = g \mathcal{P} \mathcal{L}(t) \left[1 - \Sigma(t)\right]^{-1} \mathcal{P} \equiv \sum_n g^n \mathcal{K}_{n}(t),
\end{aligned}
\end{equation}
where $\mathcal{K}(t)$ is a time-local generator, commonly referred to as the TCL (time-convolutionless) generator. The term $\mathcal{K}_n(t)$ represents the $n$th-order contribution in its perturbative expansion. The $\Sigma(t)$ in the middle term is also a superoperator, defined as
  \begin{equation}\label{eq:sigma}
\begin{aligned}
 \Sigma(t) = g\int_{t_{0}}^{t}ds\mathfrak{\mathcal{G}}(t,s)\mathcal{Q}\mathcal{L}(s)\mathfrak{\mathcal{P}}G(t,s),
 \end{aligned}
\end{equation}
with the composite system's propagator $\mathcal{G}(t,s)$ and the backward propagator $G(t,s)$:
  \begin{equation}\label{eq:propagator}
\begin{aligned}
&\mathcal{G}(t,s)\equiv T_{\leftarrow}\exp[g\int_{s}^{t}ds'\mathcal{Q}\mathcal{L}(s')],\\
&G(t,s)\equiv T_{\rightarrow}\exp[-g\int_{s}^{t}ds'\mathcal{L}(s')],\\
 \end{aligned}
\end{equation}
where $T_{\leftarrow}$ and $T_{\rightarrow}$ indicate chronological time ordering and the antichronological time-ordering. 

The above is the TCL method to get the density matrix of an open system interacting with bath, due to it's derivation in Dirac interaction picture, it only require the total system is a closed system. Generally, for the initial state of the total system is a product state $\rho(t_0)=\rho_s(t_0)\otimes\rho_b(t_0)$, one has $\mathcal{Q}\rho(t_{0}) = 0$, making $\mathcal{I}(t)\mathcal{Q}\rho(t_{0})=0$ in Eq.~\eqref{eq:tclfome}. Therefore, the evolution of \(\mathfrak{\mathcal{P}}\rho(t)\) can be obtained by calculating \(\mathcal{K}(t)\) order by order.


The first-order contribution of $\mathcal{K}(t)$ is given by $\mathcal{K}_1(t) = \mathcal{PL}(t)\mathcal{P}$. And we can always chose $\rho_{b}$ satisfies $\tr_{\mathrm{b}}[H_{i}(t)\rho_{b}] = 0$, which leads to $\mathcal{K}_{1}(t)=0$. Consequently, the lowest-order non-zero contribution of $\mathcal{K}(t)$  is  $\mathcal{K}_{2}(t)=\int_{t_0}^{t}dt_{1}\mathcal{PL}(t)\mathcal{L}(t_{1})\mathcal{P}$. Substituting $\mathcal{K}_2$ into Eq.~\eqref{eq:tclfome}, one obtains the second-order TCL equation  for $\rho_{s}(t)$:
\begin{equation}\label{eq:ctcl2}
\frac{\partial}{\partial t}\rho_{s}(t) = -g^{2}\int_{t_0}^{t} dt_{1} \tr_{\mathrm{b}}\left\{\left[H_{i}(t), [H_{i}(t_{1}), \rho_{s}(t) \otimes \rho_{b}]\right]\vphantom{\frac{9}{2}}\right\}.
\end{equation}
This is the  second order TCL equation. It's similar to the Born approximation master equation, but it's a time local equation rather than a convolution equation.

\section{Projection Operator Method with Open-system Interaction Picture}\label{sec22}
In many realistic situations, not only the target system but also its environment may be subject to external noise or dissipation. In such cases, the standard projection operator method based on a closed total system becomes inapplicable. To address this, we develop the Open Time-Convolutionless (Open-TCL) approach, which extends the projection operator method to an open-system interaction picture. By formulating a modified time-convolutionless master equation in this setting, the Open-TCL approach allows us to analytically describe the reduced dynamics of a probe system interacting with an open environment. This generalization provides an effective framework to distinguish different types of quantum dynamics via their environmental signatures.

 If the bath is additionally subjected to dissipations, the dynamics of the total system  should be governed by the Lindblad master equation~\cite{
lindblad}
 \begin{equation}\label{eq:lme}
\begin{aligned}
\frac{\partial\rho(t)}{\partial t}=-i[H,\rho(t)]+\gamma\sum_{j}\left[L_{j}\rho(t) L_{j}^{\dagger}-\frac{1}{2}\left\{L_{j}^{\dagger}L_{j},\rho(t)\right\}\right],
\end{aligned}
\end{equation}
where $L_{j}$ denotes the jump operators acting on the bath $\mathrm{B}$. Since the total system is now an open one, the traditional Dirac interaction picture cannot be employed and an interaction Hamiltonian cannot be used to generate the evolution. In order to formulate a perturbation theory for such case,  a generalization of  the perturbation theory based on the Dirac interaction picture is needed. To this end, we introduce the  {\em open-system interaction picture}, in which the density matrix is defined as 
 \begin{equation}\label{eq:lme}
\begin{aligned}
 \tilde{\rho}(t)=e^{-(\mathcal{L}_{s}+\mathcal{L}_{b})t}\rho(t),
\end{aligned}
\end{equation}
which means backward evolving $\rho(t)$ for a time duration $t$, but the Liouvillian for the backward evolution only contains $\mathcal{L}_{s}$ and $\mathcal{L}_{b}$. The evolution of $ \tilde{\rho}(t)$ is then given by
 \begin{equation}
\begin{aligned}
\frac{\partial}{\partial t}\tilde{\rho}(t)=\tilde{\mathcal{L}}(t)\tilde{\rho}(t)=e^{-(\mathcal{L}_{s}+\mathcal{L}_{b})t}\mathcal{L}_{I}e^{(\mathcal{L}_{s}+\mathcal{L}_{b})t}\tilde{\rho}(t),
\end{aligned}
\end{equation}
where 
 \begin{equation}\label{eq:}
\begin{aligned}
\mathcal{L}_{s}\tilde{\rho}(t)&=-i\left[H_{s},\tilde{\rho}(t)\right], \\
\mathcal{L}_{i}\tilde{\rho}(t)&=-i\left[H_{i},\tilde{\rho}(t)\right],\\
\mathcal{L}_{b}\tilde{\rho}(t)&=-i\left[H_{b},\tilde{\rho}(t)\right]\\
&+\gamma\sum_{j}\left[L_{j}\tilde{\rho}(t) L_{j}^{\dagger}-\frac{1}{2}\left\{L_{j}^{\dagger}L_{j},\tilde{\rho}(t)\right\}\right].\\
\end{aligned}
\end{equation}
 We also define a superoperator  $\mathfrak{\mathcal{P}}$ to project the density matrix, that is, $\mathcal{P}\tilde{\rho}(t)=\tr_{\mathrm{b}}[\tilde{\rho}(t)]\otimes\rho_{b}$. 

\section{ Application to a Qubit Coupled to a Dephasing XX Chain }\label{secxx}
With the theoretical framework established in the previous section, we now consider a specific model system. Our goal is to explore how different types of quantum dynamics can be distinguished through differences in Markovianity and non-Markovianity, by employing an auxiliary measurement system. To minimize the resources required for measurement, we choose the auxiliary system \( S \) to be a single qubit, while the target system \( B \) is taken to be an XX spin chain of \( N \) lattice sites. The qubit \( S \) is coupled to the first site of the XX chain. To better mimic realistic experimental conditions, we also introduce dephasing noise into the model. The schematic of the system is shown in Fig.~\ref{Fig:setup}.
The Hamiltonians for the qubit \( S \) (\( H_S \)), the XX spin chain \( B \) (\( H_B \)), and their interaction (\( H_I \)) are given by:
\begin{equation}
H_{s}=\omega_{0}\sigma^{+}\sigma^{-},
\end{equation}
\begin{equation}\label{eqxx}
\begin{aligned}
H_{b}=\sum_{j=1}^{N-1}(S_{j}^{x}S_{j+1}^{x}+S_{j}^{y}S_{j+1}^{y})+\sum_{j=1}^Nh_{j}S_{j}^{z},
\end{aligned}
\end{equation}
\begin{equation}\label{eqhi}
\begin{aligned}
H_{i}=g(\sigma^{+}S_{1}^{-}+\sigma^{-}S_{1}^{+}),
 \end{aligned}
\end{equation}
where $h_j$ is selected uniformly from $[-w,w]$ with $w$ being the disorder strength,  $\sigma^{\pm}=(\sigma^x\pm i\sigma^y)/2$ with $\sigma^{x/y}$ being Pauli matrices, $S^i$ ($i=x,y,z$) are spin operators, and $S^{\pm}=S^x\pm iS^y$. Here, we couples the qubit to the first site of the $\mathrm{XX}$ chain. The initial states of $\mathrm{S}$ and $\mathrm{B}$ can be simply chosen as   $|\phi(t_{0})\rangle_{s}=|1\rangle_{s}$ and $|\phi(t_{0})\rangle_{b}=|\psi_{b}\rangle=|00\cdots0\rangle_b$, respectively. Those choices make our setup quite convenient for practical realizations.  The dynamics of the single qubit will be different due to various  dynamics of the $\mathrm{XX}$ chain. Therefore, one is possible to extract dynamical information of the $\mathrm{XX}$ chain solely by measuring the single qubit. Such a non-demolition measurement will be rather convenient for practical implementations, especially when one simulating quantum systems (those target systems are chosen as the bath $\mathrm{B}$) through quantum computers.  Since the target system is now a bath for the single qubit, the dynamical information will present as Markovianity and non-Markovianity when focuing on the single qubit.

\subsection {XX Chain without Dephasing}

We first consider the closed case in which  the $\mathrm{XX}$ chain is without dephasing. Then, the evolution of the total system is determined by the Schr\"{o}dinger equation or the von Neumann equation, and our open-system interaction picture reduces to the Dirac interaction picture. Since we would like to detect the dynamics of the chain through S, we need to trace out the chain and get an effective dynamical equation for the qubit. This can be done through the TCL Method in Dirac interaction picture. In this case,
\begin{equation}
H_{i}(t) = e^{i\omega_{0}t} \sigma^{+} S_{1}^{-}(t) + e^{-i\omega_{0}t} \sigma^{-} S_{1}^{+}(t).
\end{equation}
Substituting into Eq.~\eqref{eq:ctcl2}, one obtains the second-order equation for the qubit $\mathrm{S}$:
\begin{equation}\label{eq6}
\begin{aligned}
&\frac{\partial}{\partial t}\rho_{s}(t)=-g^{2}\int_{t_0}^{t}dt_{1}\\
&\times\left\{e^{i\omega_{0}(t-t_{1})}\tr_{\mathrm{b}}\left[S_{1}^{-}(t)S_{1}^{+}(t_{1})\rho_{b}\right]\left[\sigma^{+}\sigma^{-}\rho_{s}(t)-\sigma^{-}\rho_{s}(t)\sigma^{+}\right]\right.\\
&\left.+e^{-i\omega_{0}(t-t_{1})}\tr_{\mathrm{b}}\left[S_{1}^{-}(t_{1})S_{1}^{+}(t)\rho_{b}\right]\left[\rho_{s}(t)\sigma^{+}\sigma^{-}-\sigma^{-}\rho_{s}(t)\sigma^{+}\right]\right\}.\\
\end{aligned}
\end{equation}
According to the initial state of the qubit and the qubit-chain interaction, one just needs to measure the first diagonal element of the qubit's reduced density matrix, which gives the information about the particle number in the excited level. Thus, one only needs to focus on the equation of motion for the diagonal element, and has  
 \begin{equation}\label{eq7}
\begin{aligned}
&\dot{\rho_{s}}^{11}(t)\\
=&-2g^{2}\int_{t_0}^{t}dt_{1}e^{i\omega_{0}(t-t_{1})}
\Re\left\{\tr_{\mathrm{b}}[S_{1}^{-}(t)S_{1}^{+}(t_{1})\rho_{b}]\right\}\rho_{s}^{11}(t),\\
\end{aligned}
\end{equation}
 where $\Re$ denotes the real parts. Note that in the absence of dissipations and up to second order of $g$, the equation of motion for the particle number is determined solely by the spin correlation functions of the $\mathrm{XX}$ chain, the $\tr_{\mathrm{b}}[S_{1}^{-}(t)S_{1}^{+}(t_{1})\rho_{b}]$ .  Since different dynamics of the chain will provide different spin correlations, one is possible to extract the information of dynamics through the qubit.

\begin{figure}[t]
\centering
\includegraphics[height=4.5cm]{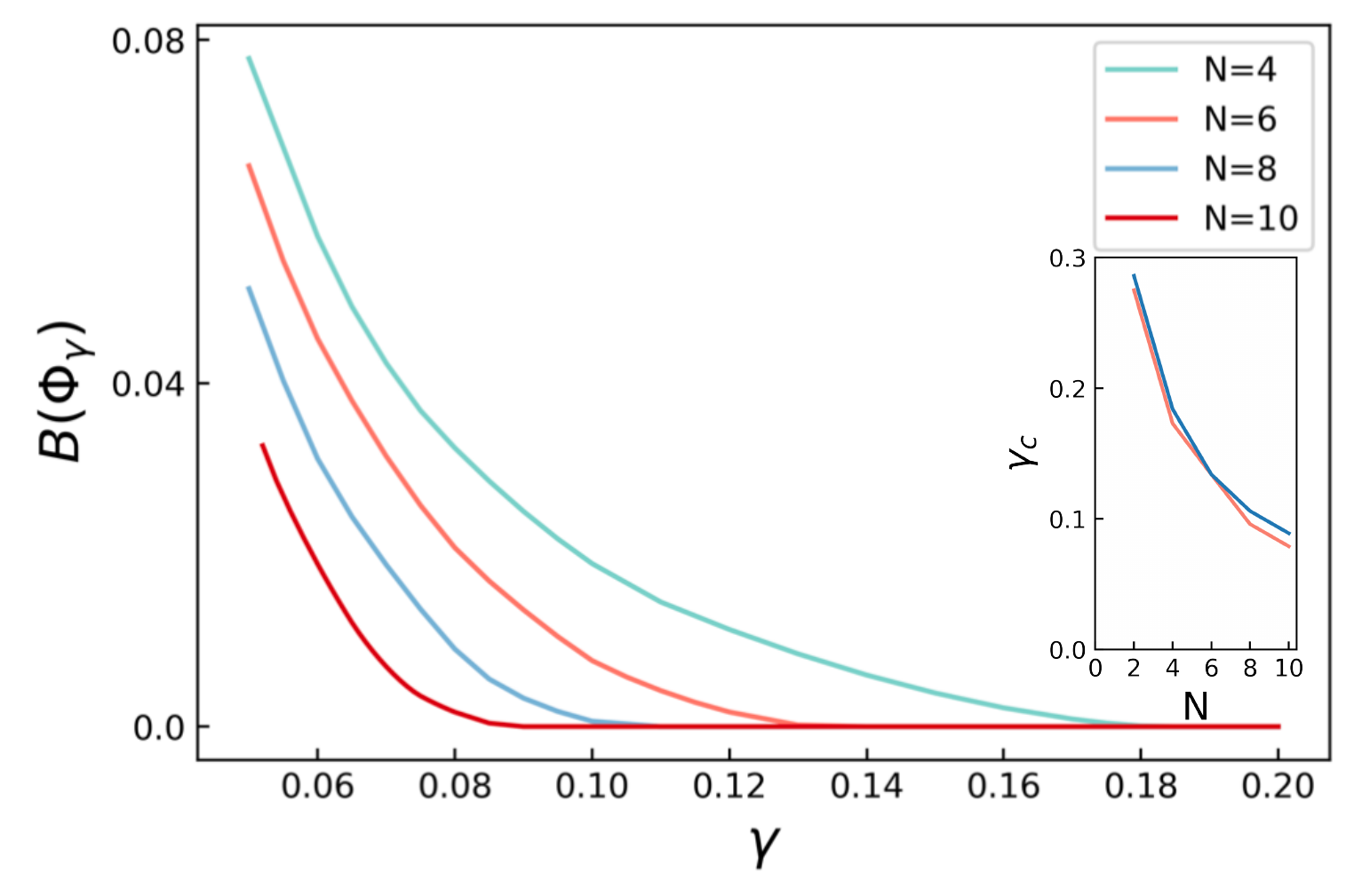}
\caption{Bounce function $B(\Phi)$ as a function of $\gamma$ for $N=4, 6, 8, 10$. Inset: $\gamma_0$ predicted by TCL2 and exact diagonalization, where the blue line shows the result of exact diagonalization, and the red line  the result of TCL2.}
\label{Fig:bounce}
\end{figure}

\begin{figure}[t]
\centering
\includegraphics[height=4.5cm]{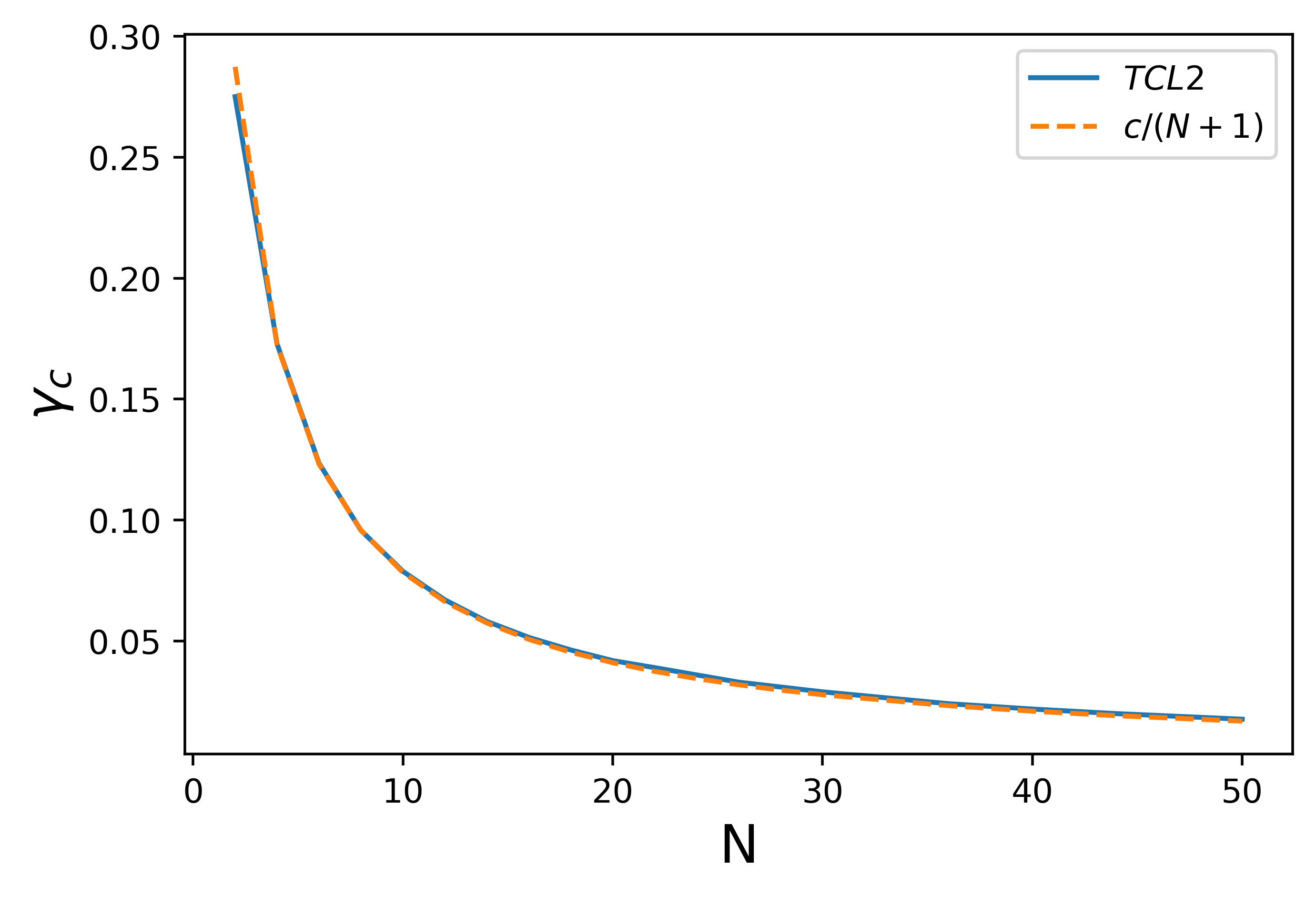}
\caption{A comparison of the critical dephasing strength $\gamma_c$ in Eq.\eqref{eq12}  as a function of $N$ (blue solid line) and the simple expression $c/(N+1)$ with $c=0.862$ (orange dashed line).}
\label{Fig:gamma_c}
\end{figure}

In the absence of disorder in the $\mathrm{XX}$ chain, the time-dependent spin correlation function can be analytically calculated by transforming the $\mathrm{XX}$ spin chain into a free fermion model via the Jordan-Wigner transformation
. The transformation is given by:
\begin{align}
S_{j}^{+} &= \exp \left( i\pi \sum_{l=1}^{j-1} \left( S_{l}^{z} + \frac{1}{2} \right) \right) a_j^{\dagger}, \\
S_{j}^{-} &= a_j \exp \left( i\pi \sum_{l=1}^{j-1} \left( S_{l}^{z} + \frac{1}{2} \right) \right).
\end{align}
Through this procedure, the Hamiltonian $H_s$ is transformed into
\begin{equation}\label{eq7_1}
H_s = \frac{1}{2} \sum_{j=1}^{N-1} J \left( a_{j}^{\dagger} a_{j+1} + a_j a_{j+1}^{\dagger} \right),
\end{equation}

where \( a_j \) and \( a_j^{\dagger} \) are fermionic annihilation and creation operators, respectively. The Hamiltonian can further be diagonalized as
\begin{equation}
H_s = J \sum_{k} \cos(k) a_{k}^{\dagger} a_{k},
\end{equation}
where
\begin{equation}
a_{k} = \left[ \frac{2}{N+1} \right]^{-1/2} \sum_{j=1}^{N} \sin(kj) a_j,
\end{equation}
and \( k = m\pi / (N+1) \) with \( m = 1, 2, \ldots, N \). Therefore, the time-dependent correlation function is
\begin{equation}\label{eq7_3}
\tr_{\mathrm{b}} \left[ S_{1}^{-}(t) S_{1}^{+}(t_1) \rho_{b} \right] = \frac{2}{N+1} \sum_k \sin^2(k) e^{-i (t-t_1) J \cos(k)}.
\end{equation}
Thus, we obtain
\begin{equation}\label{eq_8}
{\rho_s}^{11}(t) = e^{ \frac{4g^2}{N+1} \sum_k \frac{\sin^2(k) \left[ \cos \left( (\epsilon_k + \omega_0) t \right) - 1 \right]}{ (\epsilon_k + \omega_0)^{2}}},
\end{equation}
where \( \epsilon_k = J \cos(k) \). Returning to the Schr\"{o}dinger picture, the particle number in qubit \( \mathrm{S} \) is given by \( \rho_s^{11}(t) \). 

\begin{figure}[t]
\centering
\includegraphics[height=4.4cm]{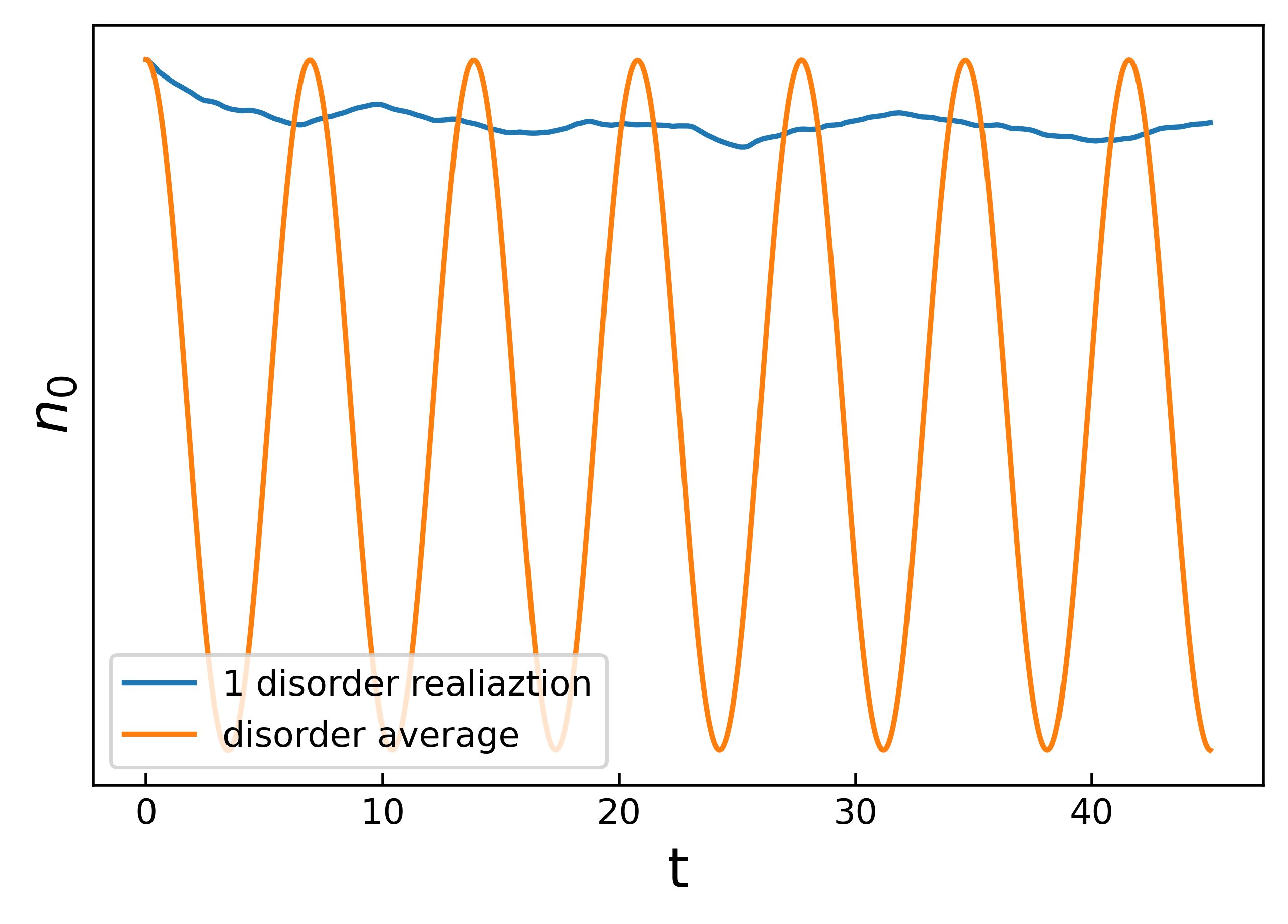}
\caption{Time evolution of the particle number in qubit ($n_0$) for one disorder realization and the resultant particle number after disorder averaging, where $W=10$.}
\label{Fig:disorder}
\end{figure}

In the localization case, we set $h_{i}\in[-W,W]$, where $W$ is the disorder strength. The finite chain will be localized \cite{Anderson:1958uc,abou1973selfconsistent} when $W$ is large. The evolution in one disorder realization is determined by Eq. \eqref{eq7}, and the disorder average time derivative of the particle number should be
\begin{equation}\label{eq15}
\begin{aligned}
\overline{\dot{\rho}_{S}^{11}(t)}=-2g^{2}\overline{\int_{t_0}^{t}dt\Re\left\{\tr_{B}\left[S_{1}^{-}(t)S_{1}^{+}(t_1)\rho_{B}\right]\right\}\rho_{S}^{11}(t)},
\end{aligned}
\end{equation}
where the over-bar denotes the disorder average. In each disorder realization, $\tr_{\mathrm{b}}[S_{1}^{-}(t)S_{1}^{+}(t_{1})\rho_{b}]$ should be an oscillation term, with the frequency of oscillation determined by $h_i$. Thus, in each realization, the dynamics of the qubit should be non-Markovian, as shown in Fig. \ref{Fig:disorder}, where the particle number in the qubit oscillates rapidly. However, after disorder averaging, this term tends to zero, making $\dot{\rho}_{s}^{11}(t)$ nearly zero.

\begin{figure}[t]
\centering
\includegraphics[height=4.4cm]{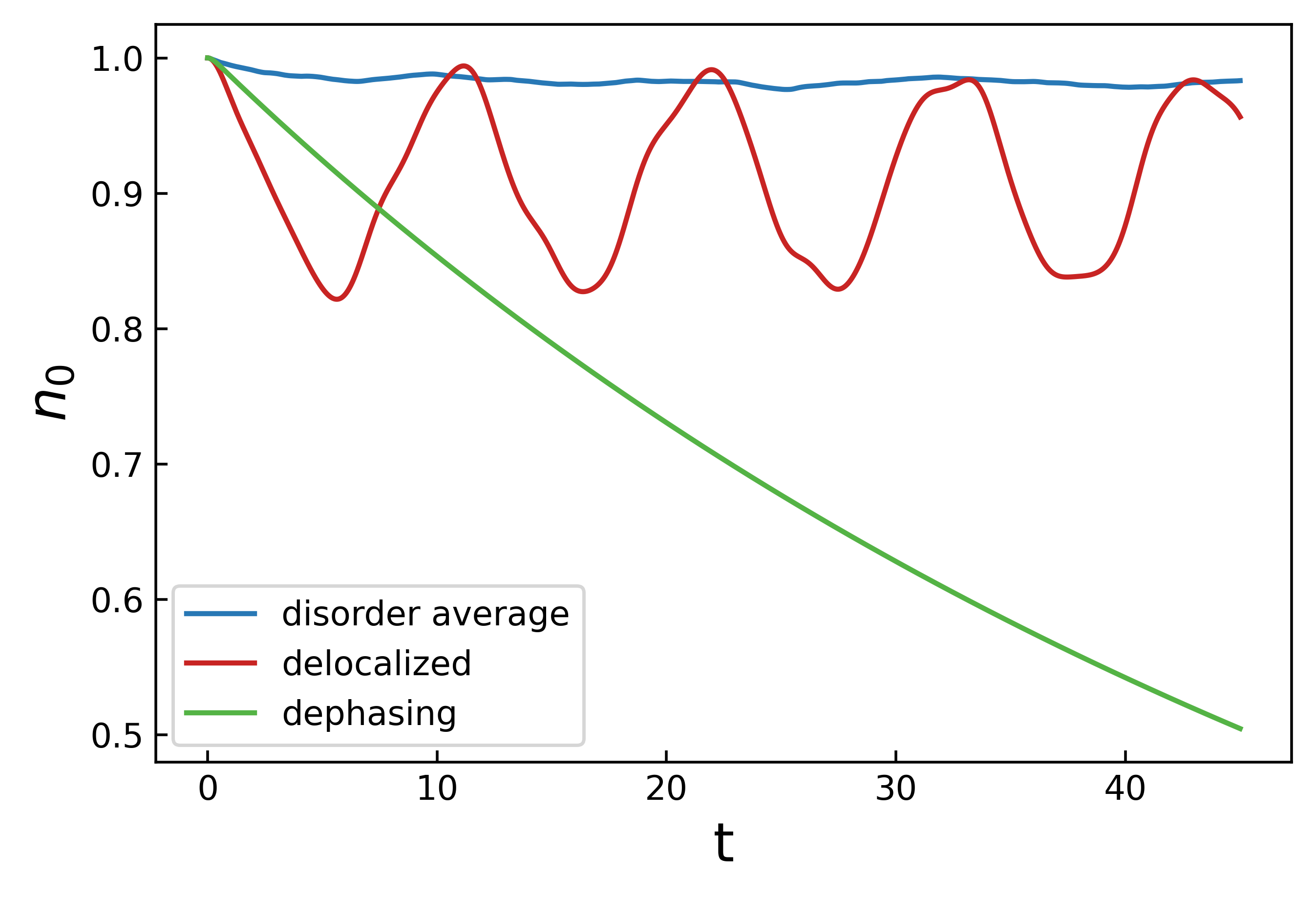}
\caption{Particle number in qubit corresponding to the different cases for $N=10$, $g=0.2$, and $\omega_0=0$, $W=10$ and $\gamma=0$ for the localized case after disorder averaging; $W=0$ and $\gamma=0$ for the delocalized case; and $W=0$ and $\gamma=2$ for the dephasing case. }
\label{Fig:distinguish}
\end{figure}

The particle number in the qubit corresponding to different cases is shown in Fig. \ref{Fig:distinguish}. This allows us to distinguish the dynamics in the chain by measuring the qubit only. For these different dynamics, the bath corresponds to either a Markovian or non-Markovian environment.


\begin{figure}[t]
\centering
\includegraphics[height=4.5cm]{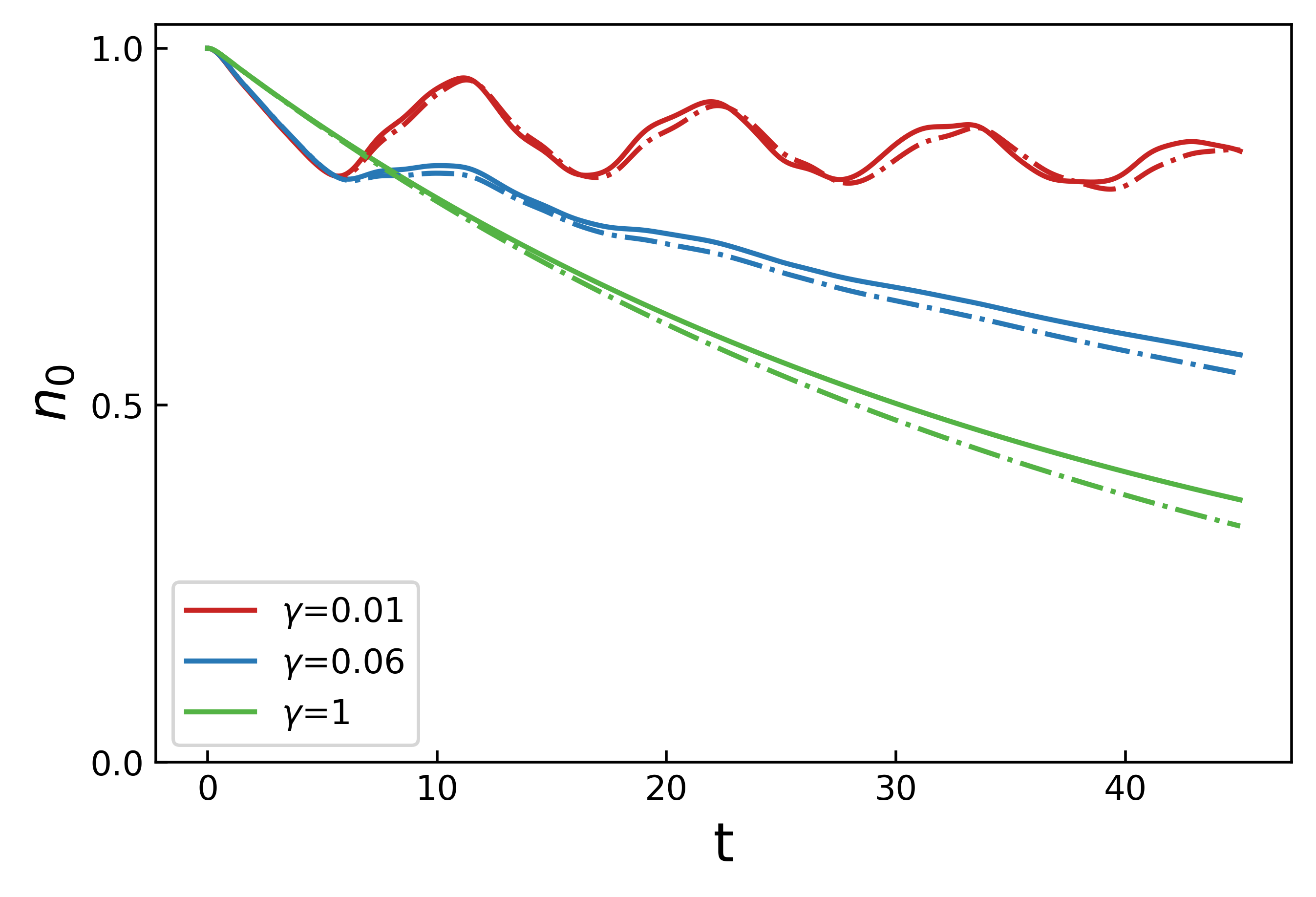}
\caption{The particle number in qubit ($n_0$) as a function of time $t$. The solid line shows the result of exact diagonlization, and the dash line the result of TCL2. The plots are made by setting $g=0.2$, $J=4$, $\omega_0=0$, and $h_i=0$.}
\label{Fig:tcl2_exact}
\end{figure}

\subsection{XX Chain with Dephasing Noise}


Suppose that the $\mathrm{xx}$ chain is subjected to the dephasing noise, which is frequently encountered in quantum simulations, then the dynamics of the total system is described by the Lindblad master equation Eq.~\eqref{eq:lme} with  jump operators being $L_{j}=\sigma^z_j$, and the perturbation theory should be exerted in the open-system interaction picture. 

The first order of $\tilde{\mathcal{K}}(t)$ can be expressed as $\tilde{\mathcal{K}}_1(t) = \mathcal{P}\tilde{\mathcal{L}}(t)\mathcal{P}$, and utilizing $\tr_{B}[\mathcal{L}_{B}(S_{1}^{+}\rho_{B})]=0$, we obtain $\tilde{\mathcal{K}}_{1}(t)=0$. Therefore, the lowest order non-zero  $\tilde{\mathcal{K}}_n(t)$ is  $\tilde{\mathcal{K}}_{2}(t)=\int_{0}^{t}dt_{1}\mathcal{P}\tilde{\mathcal{L}}(t)\tilde{\mathcal{L}}(t_{1})\mathcal{P}$, which leads us to the equation of motion for $\tilde{\rho}_{s}(t)$ up to second order of $g$:

\begin{equation}
\begin{aligned}
\frac{\partial}{\partial t}\tilde{\rho}_{s}(t)&=-g^{2}\int_{0}^{t}dt_{1}\times\\
&\{e^{i\omega_{0}(t-t_{1})}tr_{B}[S_{1}^{-}e^{\mathcal{L}_{B}(t-t_{1})}S_{1}^{+}\rho_{B}]\sigma^{+}\sigma^{-}\tilde{\rho}_{s}(t)-\\
&e^{i\omega_{0}(t-t_{1})}tr_{B}[S_{1}^{-}e^{\mathcal{L}_{B}(t-t_{1})}S_{1}^{+}\rho_{B}]\sigma^{-}\tilde{\rho}_{s}(t)\sigma^{+}+\\
&e^{i\omega_{0}(t_{1}-t)}tr_{B}[S_{1}^{+}e^{\mathcal{L}_{B}(t-t_{1})}\rho_{B}S_{1}^{-}]\tilde{ \rho}_{s}(t)\sigma^{+}\sigma^{-}-\\
&e^{i\omega_{0}(t_{1}-t)}tr_{B}[S_{1}^{+}e^{\mathcal{L}_{B}(t-t_{1})}\rho_{B}S_{1}^{-}] \sigma^{-}\tilde{\rho}_{s}(t)\sigma^{+}\}.\\
\end{aligned}
\end{equation}
Using $\mathcal{L}_{b}(a_{k}^{\dagger}|\psi_{b}\rangle\langle\psi_{b}|)=(-i\epsilon_{k}-2\gamma)a_{k}^{\dagger}|\psi_{b}\rangle\langle\psi_{b}|$,  which indicated that the dephasing noise suppresses the information transport. We get the explicit expression for the time derivative of the first diagonal element $\dot{\tilde{\rho}}_{s}^{11}(t)$, which corresponds to the time derivative of the particle numbers in qubit:


\begin{equation}\label{eq12}
\begin{aligned}
\tilde{\rho}_s^{11}(t) &=\exp\{ -\frac{4g^2}{N+1} \sum_k \frac{\sin^2(k) }{\left( (\epsilon_k + \omega_0)^2 + 4\gamma^2 \right)^2} \times \\
&\left\{ 2 \gamma  t \left( (\epsilon_k + \omega_0)^2 + 4\gamma^2 \right)+(\epsilon_k + \omega_0)^2 - 4\gamma^2 )  \right. \\
&\left. - 4(\epsilon_k + \omega_0)\gamma e^{-2\gamma t}\sin[(\epsilon_k + \omega_0)t] \right. \\
&\left. - \left( (\epsilon_k + \omega_0)^2 - 4\gamma^2 \right) e^{-2\gamma t}\cos[(\epsilon_k + \omega_0)t]\right\}\}
\end{aligned}
\end{equation}
Although the above derivation is performed in the open-system interaction picture, it can be shown that the particle occupation number \(n_0(t)\) on the probe qubit \(S\) is equivalent to that in the Schrödinger picture, i.e., \(n_0(t) = \tilde{\rho}_s^{11}(t)\). Therefore, the above expression provides the time evolution of the particle occupation on the probe qubit \(S\).
We have  analytically obtained the particle number ($n_0$) in the qubit  by  the TCL2  method. Additionally, we have computed it numerically via the exact diagonalization method. In Fig.~\ref{Fig:tcl2_exact}, we compare the results calculated with TCL2 and exact numeric solution for $N =10$. When the dephasing strength $\gamma$ is small, the particle in the qubit shows oscillation, which vanishes when the dephasing strength $\gamma$ surpasses a critical value $\gamma_s$. It is shown that the TCL2 method and exact solution lead to almost the same results in the time duration in Fig. \ref{Fig:tcl2_exact}. Notably, the TCL2 results capture the non-Markovianity for small $\gamma$ and Markovianity for large $\gamma$.

Within our parameter range, the oscillation can be attributed to the boundary effect, corresponding to the bounce from the right end of the chain and then back to the qubit. The oscillation reflects the non-Markovianity of the bath, comprising the chain and possible dephasing noise. It is worth noting that, for odd $N$ and $\omega_0=0$, the TCL2 results do not align well with the exact numeric solution. This discrepancy arises because the variable of the cosine function in Eq. \eqref{eq6} and Eq. \eqref{eq12} is  $\pi/2$, resulting in a zero in the denominator, which yields a non-negligible term and causes the perturbation theory to fail. This issue can be mitigated by introducing a non-zero $\omega_0$.

To quantify non-Markovianity, we define the bouncing function
\begin{equation}
B(\Phi) = \int_{\dot{n}_0(t)>0} dt\, \dot{n}_0(t),
\end{equation}
which depends on the dephasing strength \(\gamma\) and characterizes the degree of information backflow by integrating the positive time derivative of the particle number (\(\dot{n}_0(t) > 0\)). Physically, \(B(\Phi)\) quantifies the process by which particles bounce back from the XX chain to the qubit. It is analogous to the non-Markovianity measure introduced in the BLP framework~\cite{BreuerLainePiilo2009}, although with a slight difference: instead of considering all pairs of initial states, we focus on the distinguishability between a specific pair—the spin-up and spin-down states of the qubit. This choice is motivated by our goal of extracting environmental information through the particle occupation of the qubit. Moreover, since the total particle number vanishes when the qubit \(S\) is in the spin-down state, corresponding to a steady state of the total system, the backflow contribution also vanishes in this case. Consequently, the bouncing function \(B(\Phi)\) is equivalent to the BLP non-Markovianity measure evaluated for the chosen pair of initial states.

In Fig.~\ref{Fig:bounce}, we present the \(\gamma\)-dependent bouncing function \(B(\Phi)\) for different values of \(N\). For a given \(N\), \(B(\Phi)\) decreases monotonically with increasing \(\gamma\) and vanishes at a critical point \(\gamma = \gamma_c\). Moreover, the critical value \(\gamma_c\) decreases as \(N\) increases. In the limit \(\gamma \to 0\) with finite \(N\), \(B(\Phi)\) diverges. However, for any finite \(\gamma \neq 0\), the system eventually reaches a steady state, and \(B(\Phi)\) remains finite.
\begin{figure}[t]
\centering
\includegraphics[height=4.4cm]{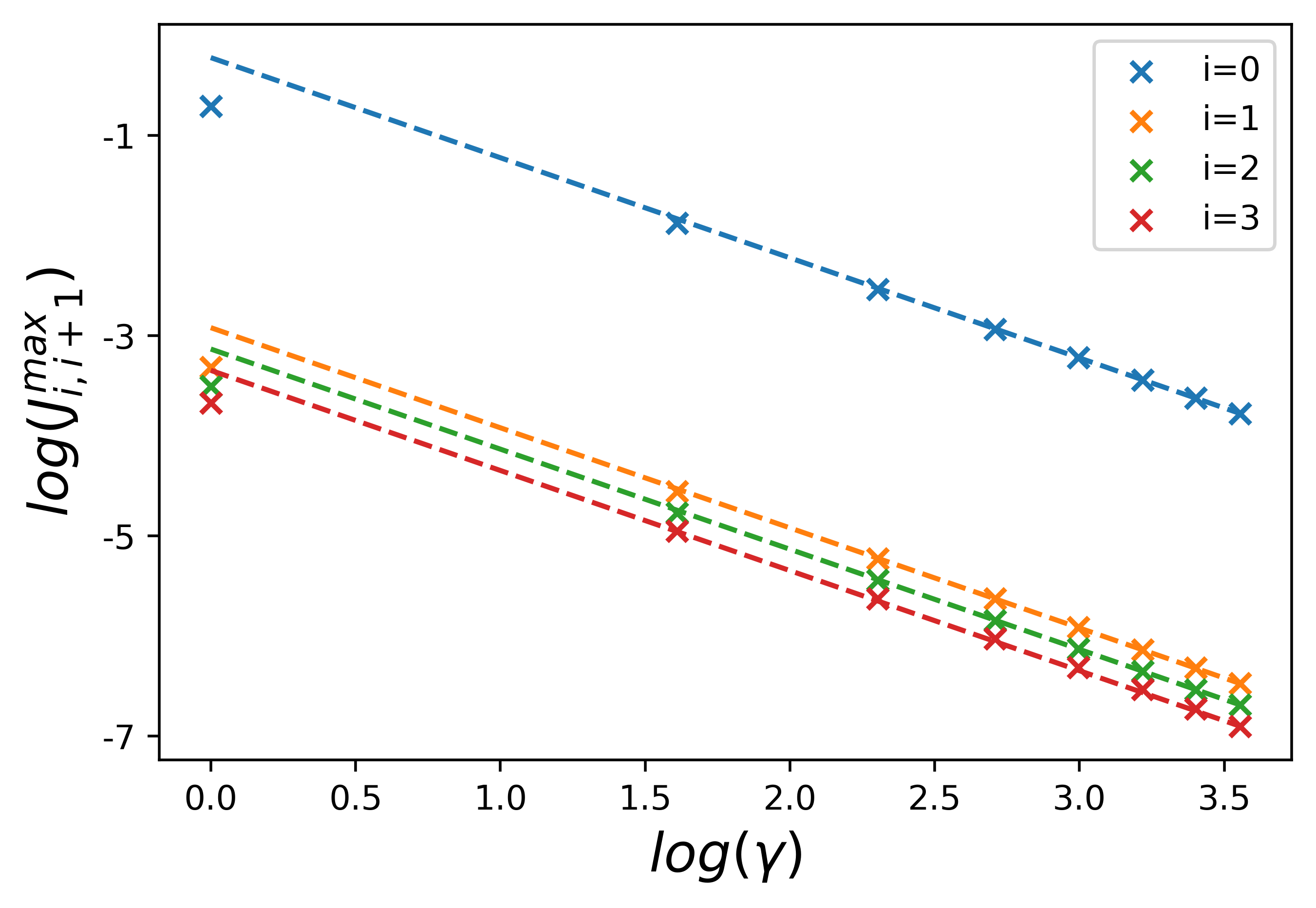}
\caption{The maximum values of the current between site $i$ and site $i+1$, denotes as $J_{i,i+1}^{max}$, plotted as a function of $\gamma$ for $i=0, 1, 2, 3$. The slope of the double log plots is equal to $-1$. }
\label{Fig:current}
\end{figure}

\begin{figure}[t]
\centering
\includegraphics[height=4.2cm]{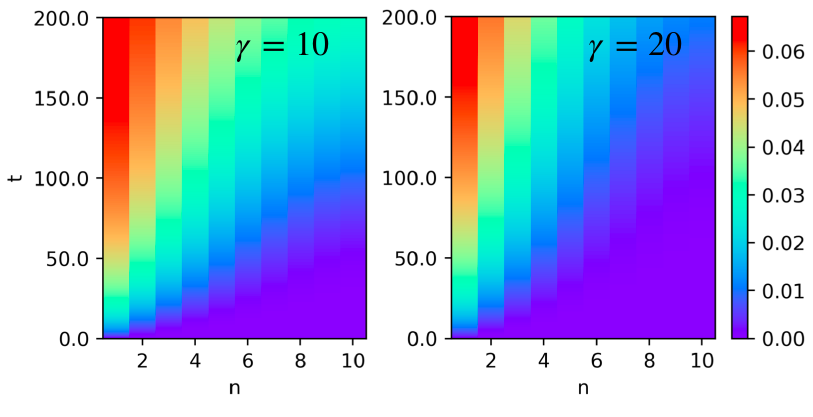}
\caption{The particle number in the chain (with the scale shown on the right side) depending on lattice point $n$ and time $t$ for $\gamma=$10 and 20, with the parameters: $g=0.2$, $J=4$, $\omega_0=0$, and $h_i=0$.}
\label{light_cone}
\end{figure}

We can derive the critical dephasing strength \(\gamma_c\) from Eq.~\eqref{eq12}. For even \(N\) and \(\omega_0 = 0\), when \(N\) becomes large, the time required for particles to return from the boundary increases, suggesting that \(\gamma_c\) should decrease as the chain lengthens. In the limit \(N \to \infty\), it is expected that \(\gamma_c\) tends to zero. In Eq.~\eqref{eq12}, the coefficient in front of the summation is close to \(\tan^2(k)\). Since the derivative of \(\tan^2(k)\) diverges near \(k = \pi/2\), the summation is dominated by modes where \(k\) is closest to \(\pi/2\). In particular, the mode with \(k = \frac{N\pi}{2(N+1)}\) provides the leading contribution.
A rough estimate of \(\gamma_c\) can be obtained by analyzing the inequality in Eq.~\eqref{eq13}:
\begin{equation}\label{eq13}
\left[e^{-2\gamma t}\sqrt{\epsilon_{k}^{2}+4\gamma^{2}}\sin\left(\epsilon_{k}t+\arctan\left(\frac{-2\gamma}{\epsilon_{k}}\right)\right)+2\gamma\right]_{\mathrm{min}} > 0,
\end{equation}
where the first term on the left-hand side represents a triangular function modulated by an exponentially decaying envelope, indicating that each mode in the XX chain is exponentially suppressed by dephasing noise. This inequality holds for \(\gamma > \gamma_c\), allowing us to approximate the critical dephasing strength as
\begin{equation}
\gamma_{c} \approx J \frac{\mathrm{ProductLog}\left(\frac{3\pi}{2}\right)}{3\pi} \sin\left(\frac{\pi}{2(N+1)}\right) \approx \frac{0.862}{N+1}.
\end{equation}
This estimate shows good agreement with the results obtained from the TCL2 master equation across all modes, as illustrated in Fig.~\ref{Fig:gamma_c}.
Moreover, since the bouncing function \(B(\Phi)\) can be directly constructed from the time evolution of the particle number on the probe qubit \(S\), the critical Markovian-to-non-Markovian transition characterized by \(1/\gamma_c\) can be inferred solely through measurements on the ancilla. This indicates that measuring the ancillary qubit allows one to track the transition between Markovian and non-Markovian behavior of the target system without requiring direct access to its full dynamics.

In the presence of strong dephasing noise, we investigate the particle transport in the chain. Our result have shown that the current $J_{i,i+1}=2(\sigma_{i}^{x}\sigma_{i+1}^{y}-\sigma_{i}^{y}\sigma_{i+1}^{x})$ is suppressed by the dephasing noise. Fig. \ref{Fig:current} showcases $\log (J_{i,i+1}^{max})$ as a function of $\gamma$. The plot of  $\log (J_{i,i+1}^{max})$ against $\log (\gamma) $ yields a straight line with a slope of -1 in the high $\gamma$ domain, signifying that the peak current value is inversely proportional to $\gamma$ in the realm of strong dephasing noise. Furthermore, the light cone of the  particle transport delays with $\gamma$ increasing, as shown in Fig. \ref{light_cone}. This is similar to the $1/\gamma$  suppression in  the diffusive coefficient of dephasing free fermion chain\cite{yang2022keldysh}.

\section{Discussion}\label{sec4}

In this work, we studied how to derive the effective dynamics of a subsystem when the total system itself is open. To this end, we introduced the open-system interaction picture and, based on this framework, generalized the existing time-convolutionless (TCL) master equation to obtain the open-TCL master equation. We then applied this method to a quantum-assisted measurement scenario, where an ancillary qubit is used to probe the nonequilibrium dynamics of a target system, with the target acting as an environment for the probe.

Specifically, we considered the target system to be an XX spin chain subject to dephasing noise, with the ancillary system being a single qubit coupled to the first site of the chain. Using the open-TCL approach, we analytically derived the master equation for the reduced dynamics of the qubit and obtained explicit expressions for the time-dependent particle number \(n_0(t)\) under different dynamical regimes of the spin chain.

We explored three distinct types of dynamics: (i) delocalized dynamics without dephasing, (ii) localized dynamics without dephasing, and (iii) diffusive dynamics under strong dephasing. For the first two cases, where the total system remains closed, we utilized the traditional interaction picture to derive the second-order master equation for the qubit. For the third case, where the total system is open due to dephasing, we employed the open-system interaction picture.

Our findings reveal that the behavior of the ancillary qubit faithfully reflects the dynamical properties of the spin chain. In the delocalized case, the qubit exhibits periodic oscillations due to boundary-induced particle bouncing, indicating non-Markovian dynamics. In the localized case, rapid strong oscillations of the particle number emerge due to the localization mechanism, leading to a distinct non-Markovian behavior that occurs much earlier than boundary-induced effects. Upon averaging over disorder realizations, the particle number fluctuations are significantly suppressed in the localized scenario, allowing a clear distinction from the delocalized case.

With the introduction of dephasing noise, the dynamics become diffusive. As the dephasing strength \(\gamma\) increases, particle bouncing is gradually suppressed, and the degree of non-Markovianity decreases. Above a critical value \(\gamma_c\), the system transitions to a Markovian regime, and the particle number in the qubit monotonically decays to its steady-state value. This transition is captured by the bouncing function \(B(\Phi)\), which vanishes when \(\gamma > \gamma_c\).

We further derived an approximate analytical expression for \(\gamma_c\) by analyzing the dominant modes in the spin chain. The critical value \(\gamma_c\) scales inversely with \(N+1\), where \(N\) is the number of sites, implying that in the thermodynamic limit, any finite dephasing strength renders the bath effectively Markovian.

These results demonstrate that by measuring only the ancillary qubit, one can efficiently infer the transition between Markovian and non-Markovian behavior of the target system without requiring direct access to the full system dynamics. This provides a resource-efficient strategy for probing the dynamical properties of complex quantum systems through indirect measurements.

\section{Conclusion}

We have developed the open-TCL master equation as a generalization of the standard TCL formalism to accommodate situations where the total system is itself open. This framework enables the analytical treatment of reduced dynamics even when the environment experiences external dissipation.

By applying this method to a qubit–XX spin chain system, we demonstrated that the reduced dynamics of the probe qubit can effectively capture key features of the environment, including the transition from non-Markovian to Markovian behavior. Our results highlight the potential of indirect quantum probing strategies and extend the applicability of TCL-based methods to a broader class of open quantum systems.

\begin{acknowledgments}
This work is supported by the Nature Science Foundation of China (Grant No.11974393) and the Strategic Priority Research Program of the Chinese Academy of Sciences (Grant No. XDB33020100). 
\end{acknowledgments}

\bibliography{KNSM_Ref}

\end{document}